\documentclass[sigconf, nonacm=true]{acmart}
\usepackage{amsmath}

\usepackage{amssymb}
\usepackage{booktabs} 

\usepackage{subcaption}
\usepackage{graphicx}
\graphicspath{ {./imgs/} }

\usepackage{url}
\usepackage{breakurl}
\usepackage{multicol}
\usepackage{fancyvrb}
\usepackage{float}
\usepackage{xcolor}
\usepackage{array}
\usepackage{multirow}
\usepackage{tabularx}
\usepackage{colortbl}
\usepackage{hhline}
\usepackage{textcomp}

\usepackage[T1]{fontenc}
\usepackage[utf8]{inputenc}
\usepackage{lmodern}

\usepackage[linesnumbered]{algorithm2e}

\setcopyright{none}

\begin{document}
\title{Automated Evaluation of Web Site Accessibility\\
 Using A Dynamic Accessibility Measurement Crawler}

\author{Trevor Bostic, Jeff Stanley, John Higgins, Daniel Chudnov, Justin F. Brunelle, Brittany Tracy}
\affiliation{
  \institution{The MITRE Corporation}
  \country{USA}
}
\email{{tbostic, jstanley, jphiggins, dlchudnov, jbrunelle, btracy}@mitre.org}

\begin{abstract}

Achieving accessibility compliance is extremely important for many government agencies and businesses who wish to improve services for their consumers. With the growing reliance on dynamic web applications many organizations are finding it difficult to implement accessibility standards, often due to the inability of current automated testing tools to test the stateful environments created by dynamic web applications. In this paper, we present mathematical foundations and theory for the Demodocus framework and prototype, and outline its approach to using web science, web crawling, and accessibility testing to automatically navigate and test interactive content for accessibility. Our approach simulates the page interactions of users with and without disabilities, and compares graphs of reachable states from these simulations to determine both the accessibility and the difficulty of content access for these different users.

\end{abstract}






\maketitle

\pagestyle{plain}

\section{Introduction}
\label{intro}

Accessibility of online content is becoming increasingly important as governments and many other organizations are moving critical functions and services online. Nearly 20\% of people in the U.S. have a disability \cite{census} and many of them are affected by inaccessible websites.

There are several barriers that are preventing web content owners from implementing sufficient accessibility practices: lack of knowledge, lack of testing capability, and a lack of funding. The lack of knowledge is being addressed in many ways, such as the development of web accessibility standards (e.g., W3C \cite{wai}) and training programs for testers and developers (i.e., Trusted Tester Program \cite{dhsttp}). The Trusted Tester Program trains workers to use automated and  manual tools to determine if applications meet the accessibility criteria. The emphasis on training and manual accessibility review is required in large part due to the current capabilities of automated accessibility tools. Current estimates suggest that automated tools are only capable of finding at most ~50\% of issues \cite{vigo2011}, though this number continually fluctuates, both up and down, due to the arms race between automated tools and the web's ever-increasing dynamism \cite{ijdl}. As a result, fixing inaccessible systems may cost a government agency or business hundreds of thousands of dollars a year, reducing their ability to add new features as they attempt to conform to accessibility standards. Accessibility testing that is reliant on manual and semi-automated testing will continue to strain funding resources as the web grows increasingly dynamic.

Web applications that rely on JavaScript and Ajax to construct dynamic web pages pose further challenges for both persons with disabilities and for automated accessibility evaluation tools. In these increasingly popular web applications, content changes in ways that cannot be easily predicted by examining a page's source code (e.g., HTML). Current automated tools can no longer discover and test the full range of web content \cite{jbrunelleDissertation}, leading to un-discoverable accessibility violations.

Our goal is to increase the capability of automated testing tools so that accessibility testers will be able to evaluate a larger percentage of accessibility problems more quickly, resulting in fewer dollars spent per page tested and a higher quality service for persons with disabilities. We present an automated testing crawler to demonstrate a reduction of effort and an improvement to the existing methods of manual and semi-automated testing that make up the current state-of-the-art.

Our approach combines concepts from web application testing, web science, and accessibility to create a methodology and prototype for automated identification of accessibility issues caused by dynamic state changes of a website. We use a web crawler to calculate and compare the difference in content reachability between users with and without disabilities. Specifically, we generate a graph of reachable states and flag errors on states that may be unreachable for some users with disabilities. Further analysis by developers and testers may be necessary to remediate violations.


In this paper, we introduce the mathematical foundations and theory underlying \emph{Demodocus}: a framework and proof-of-concept software package\footnote{Code available at https://github.com/mitre/demodocus} for automatically assessing the accessibility of websites \cite{bostic2020demodocus}. Demodocus uses web application testing theory, web crawling concepts, and web accessibility standards to perform accessibility assessments. In this paper, we discuss the mathematical model for quantifying accessibility violations based on a graph of reachable states in a state machine representing a user's interaction within an application.

\section{Related Work}
\label{relatedwork}

Three streams of research inform Demodocus. First, we discuss current web accessibility work to include the existing guiding standards for website accessibility along with existing research that explores automatically identifying and quantifying accessibility. This analysis of prior works establishes the current state-of-the-art in accessibility measurement and accessible web design.

Second, we present prior work describing web user interaction modeling techniques. The techniques provide methods for simulating web user interactions -- both manually and automatically -- to interact with and discover content from web pages. This confluence of research establishes the ways in which previous researchers have designed user interactivity models for the use of crawling and testing and, therefore, informs our development of accessibility user models.
Finally, we summarize literature from the web science domain on automatically crawling web applications (e.g., those driven by JavaScript and Ajax) for both testing and evaluation purposes. These works inform our web crawler design, including the models for state equivalency.

\subsection{Approaches for Web Accessibility Measures}
\label{accessibilitymeasure}

The need to measure or test for accessibility of websites is well established. The Web Content Accessibility Guidelines (WCAG) were designed to create an acceptable level of performance for users with disabilities by setting requirements for standard practices and principles \cite{wcag}. 

WCAG standards continue to evolve as a response to current web design trends (e.g., adoption of JavaScript to build websites) and attempt to guide the state of web practices toward increasing accessibility. Harper and Chen used the Wayback Machine\footnote{\url{http://web.archive.org/}} to conduct a longitudinal study of the accessibility of both popular and random websites using WCAG guidelines. The found that accessibility standards are met on less than 10\% of web pages, potentially due to the increased adoption of Ajax \cite{harper2012}. Additionally, Hackett et al. showed that websites are becoming less accessible over time \cite{harper2012}. Together, these studies suggest that accessibility practices are increasingly difficult to implement and sustain as the dynanism on the web increases.

To combat this difficulty, many web developers now follow a process of using automated tools for triaging violations, then following up with manual efforts that are more likely to catch subtle violations. For example, a standard accessibility reviewer might first run a fully automated tool such as Axe to quickly identify any issues found purely in the HTML and CSS \cite{axe}. They may then run pre-scripted accessibility tests on the page; the General Services Administration uses the pa11y tool \cite{pa11y} to crawl and assess the accessibility of both static HTML and the HTML that results from scripted user interactions. Finally, the reviewer will manually test the page, often using tools like the  Accessible Name and Description Inspector (ANDI) and screen readers to help diagnose accessibility issues more quickly \cite{ANDI}. Unfortunately, each step in this process has a weakness: Automated tools are fast but only catch at most 50\% of known issues, scripted tools lack the ability to discover and interact with the Document Object Model (DOM) without \textit{a priori} knowledge and human scripting of a tool, and manual tools rely on user expertise and are often very time intensive. Despite best efforts, the current testing and evaluation process remains manually burdensome and susceptible to error.

\subsection{Web User Interaction Models}
\label{usermodels}

To model users' navigation around the web, researchers have used agent-models based on varying degrees of prior knowledge. Bartoli et al. demonstrated the ability to record and replay navigations on Ajax-reliant resources \cite{navigationReplay}. Their tool records interactions and replays the activity as a method to test user interfaces with the goal of overcoming the difficulties in automated testing and programmatic interaction with client-side interfaces introduced by Ajax.

Rather than replay interactions, Gorniak worked to predict future user actions by observing past interactions of other users \cite{Gorniak:2000:PFU:647288.721746}. This work modeled users as agents based on their interactions and modeled clients (i.e., states of a representation) based on potential state transitions. The agents would make the next most probable state transition, indicating what action the user is expected to take.

Leveraging user interaction patterns, Palmieri et al. generated agents to collect hidden web pages by identifying navigation elements from current 
interactive elements \cite{autoAgents}. They represented the navigation patterns from these agents as a directed graph and trained the agents to interact with navigation elements. This includes filling out forms based on \textit{a priori} knowledge and randomly selecting options from drop-down menus to spawn new pages. This method generated  coverage that included 
80\% of all potential pages.

\subsection{JavaScript Web Application Crawling}
\label{jstesting}

Several efforts have studied client-side state exploration with attention toward testing Rich Internet Applications (RIAs; i.e., web pages that rely heavily on JavaScript to provide functionality and interactivity). This continues to be a difficult problem since RIAs often have deferred representations\footnote{\emph{Deferred representations} refer to web-based representations that rely on JavaScript to build the client-side representation, including making HTTP requests for additional embedded resources after the initial page load \cite{jbrunelleDissertation}.} that are invisible or unavailable to traditional automatic web crawlers. Rosenthal describes the web as evolving from delivering document markup as representations to delivering programs/applications as representations  \cite{iipc2013, futureWeb}. This makes reliably crawling a representation challenging, similar to capturing a PDF snapshot versus a video of a program's execution. Further exploring the difficulty of RIA state exploration, Mesbah et al. described problems they encountered, such as browser cookies and HTTP 404 responses, when trying to recreate states for a representation \cite{mesbah}.

When crawling web pages, the most common strategy is to create a graph in which events are represented as edges and states are represented as nodes. Raj et al. performed crawls on Ajax-driven RIAs and detailed a method of using state transitions defined by JavaScript events and state equivalence using the DOM trees for comparison \cite{raj2018}. Using a similar approach, Mesbah et al. performed several experiments regarding crawling and indexing representations of web pages that rely on JavaScript \cite{mesbahCrawling, mesbahInferState, mesbahDiss} focusing mainly on search engine indexing and automatic testing \cite{mesbahTesting, mesbah2}. 

Li et al. defined state transitions in the same way as Raj et al., but used a different technical approach — injecting JavaScript into crawled pages via a proxy [17]. Fejfar crawled RIA DOM elements using a human-in-the-loop approach, allowing the human to direct the user interactions that should be crawled \cite{fejfar2018} (similar to pa11y \cite{pa11y}).

The primary basis for our work is the research performed by Dincturk et al. \cite{dincturkAjax, mappingState, dincturkDiss}. They present a model for crawling RIAs by constructing a graph of client-side states (they refer to these as ``AJAX states'' within the Hypercube model). Their work identifies the challenges with crawling Ajax-based representations and uses a Hypercube strategy to efficiently identify and navigate all client-side states of a deferred representation. Their model defines a client-side state as a state reachable from a URL through client-side events and is uniquely identified by the state’s DOM. That is, two states are identified as equivalent if their DOM is directly equivalent.

\section{Demodocus Design}
\label{design}

As discussed in Section \ref{intro}, our work resulted in a framework we refer to as Demodocus for measuring accessibility that leverages principles from web science, web accessibility, and -- specifically -- automated web application testing. The resulting software package includes a web crawler, a set of user models, and a method by which states are assessed for equivalence and accessibility violations.

Demodocus (described in Section 3.1), takes a Uniform Resource Identifier (URI) as input and creates a graph of the reachable states within the webpage. Demodocus exercises user interactions and client-side events to reach the states. As Demodocus creates the graph, it uses a state equivalency algorithm (discussed in Section \ref{models}) to identify and deduplicate potentially equivalent states. Deduplicating states within the built graph reduces the state space to a reasonable level presentable to an end user (e.g., tester or developer) and prevents duplicated states from skewing or diluting accessibility results.

The canonical graph is created using an omnipotent user model (the \textit{omni} user) capable of finding and triggering all possible events. This results in a complete graph of the states available through interaction with the page. To assess accessibility (Section \ref{inaccessible}), additional user models based on the capabilities of persons with and without disabilities then traverse through the complete graph. While traversing, Demodocus records the subgraph of states each user model is able to reach. 

After completing the crawl, we compare the user models' graphs of reachable states and access paths for each state (i.e., the series of events and states a user must traverse to reach an end state) to the complete graph generated by the omniscient user. By finding the differential between the graphs, we can ascertain which states were unreachable (and therefore potential accessibility violations) for a given user model. Additionally, we can gain insight on the difficulty of traversal to some end state (i.e., the ease or difficulty of reaching a portion of the web application) as compared to the optimal path (e.g., the omniscient user may have a significantly shorter access path than a given user model). In a fully accessible web page, we would expect the differential between the complete graph and the graphs generated by the user models to be minimal.

The user models (described in Section 3.3) are specifically made to simulate the abilities of persons with and without disabilities as they navigate a webpage  (i.e., traverse the graph). The models in this paper simulate only a small set of disabilities that a user may have (e.g., the inability to effectively use a mouse to trigger client-side events such as mouseover), but can provide adequate insight to predict poor experiences a user may face while using a web page. We expect that these models will be significantly adapted and expanded in the future.

\subsection{Algorithm}
\label{crawler}

Demodocus uses a similar methodology to our prior research focused on web archiving \cite{brunellejcdl2018}. Using Selenium \cite{seleniumpjs}, the crawler begins by dereferencing a URI provided as a parameter to the crawler application. Demodocus then creates a frontier of available user interactions and -- similar to a traditional web crawler -- executes the events on the client. Demodocus then records the newly reached states. Because modern web sites more closely resemble web \emph{applications} (i.e., RIAs) through the use of JavaScript and other client-side events to generate content or load embedded resources, the crawler uses JavaScript utilities to exercise and monitor the client-side activity; this is a capability not available to traditional web crawlers (such as Heritrix \cite{heritrixProductionVersion}).

\subsubsection{State Machine Definition}

Within our crawler model, we first use a breadth-first search model to create a complete graph of all states reachable by client-side user interactions. We represent the generated graph abstractly as a state machine $M$, defined in Equation \ref{machinem}.


\begin{equation}
\begin{split}
M &= (S, s_0, \Sigma, \delta)\\
S &= \text{All client-side states}\\
s_0 &= \text{Representation available after}\\
&  \text{dereferencing the URI}\\
\Sigma &= \text{All client-side interactions}\\
\delta&: \ S \times \Sigma \rightarrow S\\
s_i, s_j &\in S\\
\sigma &\in \Sigma \\
\delta\text{($s_i$, $\sigma$)} &= s_j\\
\end{split}
\label{machinem}
\end{equation}

Our definition of $S$ includes all states possibly found, even those found by waiting (e.g., performing no action until a timeout event occurs) and does not take into account that human users may consider the states to be duplicate or equivalent. This means we could have states $s_i$ and $s_j$ that have equivalent content but may be considered separate states for other reasons (such as the paths to reach them). We have also defined $s_0$, the initial state, as the state available after dereferencing the URI. 

Additionally, we define our alphabet $\Sigma$ as the set of client-side interactions that lead to states in $S$. Note that $\Sigma$ includes client-side interactions such as waiting and should not be assumed to be a set of JavaScript events performed on some element.

Finally, we define a transition function $\delta$ that maps a state and an interaction to an output state. We will use this state machine as a model to help define accessibility violations in the context of our graph analysis.

\subsubsection{Graph Implementation}

In practice, it is impossible to fully exercise the state machine defined above, since there may be infinite states due to our consideration of interactions such as waiting or page reloading. Thus our implementation follows an approximation of the state machine that begins by generating a graph $G$ (defined in Equation \ref{graphg}) of the states and interactions crawled. 



\begin{equation}
\begin{split}
G &= (V, E)\\
V &= \text{Representation states reachable}\\
  &  \text{via user interaction}\\
E &= \text{User interactions (transitions)}\\
\end{split}
\label{graphg}
\end{equation}


Within our graph $G$, we attempt to reduce the state search space (i.e., reduce $|V|$) to increase tractability of our crawl. We collapse similar states, as defined by our State Equivalence model (Section 3.3), into one vertex within our graph. That is, there may exist states $s_i$ and $s_j$ within our state machine $M$ that are collapsed to $v_i$ within our graph $G$. Note that this implies the canonical graph $G$ is fully dependent on our State Equivalence model.  

Additionally, each vertex $v$ in our graph contains a set of some targets  $T_{v} \in T$ with which our crawler can interact, where $T$ is the set of all possible targets available throughout the graph. $T$ is dependent on the representation, as it is possible that a target $t$ is available only in one or more representations $v_i$ and $v_j$. For ease, we denote the set of targets available in a given representation $v_i$ as $T_{v_i}$. Examples of possible targets may include interactive web elements or the browser window.

Each edge $e \in E$, representing a user interaction, is a pair of some action $a \in A$ and a target $t \in T_{v_i}$ where action $a$ is performed on target $t$ within representation $v_i$ to trigger the transition to $v_j$. 

$A$ is the set of all available actions and is independent of the representation. That is, $A$ is a set of all actions users may attempt at any time. For example, actions could include clicking, keyboard presses, or reloading the window.  

The full definition of how states and edges are represented in our graph is given in Equation \ref{edge_def}. 

\begin{equation}
\begin{split}
G &= V,E \\
A &= \text{Set of available actions} \\
T &= \text{Set of possible targets} \\
v_i &\in V \text{ State in the graph}\\
T_{v_i} &= \text{Set of targets available in $v_i$} \\
a &\in A \text{ Action available to user} \\
t &\in T \text{ Target element capable of interaction} \\
e &\in E \text{ Edge in the graph} \\
e &= \{a, t\} \text{ Edge represent an action, target tuple} \\
\end{split}
\label{edge_def}
\end{equation}

\subsubsection{User Model Dependent Graphs}
To create our graphs we employ a variety of user models that are meant to simulate the experience of persons with and without disabilities. Thus, each may generate a different graph $G$ that represents the reachable states $V$ and executable interactions for that user $E$ (e.g., $G_{omni}$, $G_{non-disabled}$, $G_{keyboard}$, $G_{screen-reader}$). An example of our notation for this structure with omni-user is shown below in Equation \ref{omni}. 


\begin{equation}
\begin{split}
G_{omni} &= (V_{omni}, E_{omni})\\
V_{omni} &= \text{All states reachable by omni-user}\\
E_{omni} &= \text{All interactions executable by omni-user}\\
\end{split}
\label{omni}
\end{equation}
 

Among the user models, the omni-user model has no ability or navigation restrictions placed upon it. Due to this, we utilize omni-user to create our complete graph $G_{omni}$ (or a graph as complete as is feasible), defined in Equation \ref{omni}, to act as ground truth for the remaining user models. We then crawl the complete graph with the restricted user models to determine which states are reachable and which edges are traversable. For clarity, it should be noted that the omni-user is meant only for graph creation purposes and is not modeled after any realistic user. Subsequently, the restricted users \emph{are} modeled after real users and contain models for users with and without disabilities.

Our restricted user models $r$ are each defined with a set of actions $A_r$ they are capable of performing, as well as a set of targets $T_r$ on which they are capable of performing actions. These sets are restricted by the capabilities of the user they are simulating. For instance, a keyboard user would not have any mouse events within their set of actions $A$, a color blind user would have a reduced set of targets $T$ based on the target's color contrast, and a user without a disability would similarly have no invisible elements in its set of targets. 

Each restricted user model will generate its own graph $G_r$ containing the reachable representations $V_r$ and the interactions they were able to execute $E_r$. The graph $G_r$ is a subset of the graph $G_{omni}$; however, the restricted users may be able to find all representations $V_{omni}$ and execute all interactions $E_{omni}$, making $G_r$ a potential improper subset, thus $G_r \subseteq G_{omni}$.

These properties of our restricted user models are stated concisely in Equation \eqref{user_model_properties}.

\begin{equation}
\begin{split}
G_{omni} &= V_{omni}, E_{omni} \\
E_{omni} &= \{A_{omni}, T_{omni}\} \\
A_r &= \text{Set of action available to a restricted user} \\
T_r &= \text{Set of targets available to a restricted user} \\
A_r &\subseteq A_{omni} ,\; T_r \subseteq T_{omni} \\
E_r &= \{A_r, T_r\} \\
& \text{so} \\
V_r &\subseteq V_{omni} ,\; E_r \subseteq E_{omni} \\
G_r &= V_r, E_r \\
& \text{thus}  \\
G_r &\subseteq G_{omni} \\
\end{split}
\label{user_model_properties}
\end{equation}

Remember, restricted user models encompass users with and without disabilities, as neither has access to the full spectrum of capabilities available to the omni-user model.




\subsubsection{Finding Inaccessible Representations}
\label{inaccessible}

Within the context of state machine $M$ and graph $G$, we define an inaccessible representation to occur when there exists some $v \in V_{omni}$ with corresponding incoming edge $e_x = \{a_x, t_x\}$ that is available to non-disabled users $n$, but there is no available incoming edge $e_y = \{a_y, t_y\}$ for users $d$ with a disability. Restated, this means that an incoming edge to state $v$ exists where for at least one of the incoming edges $e_x$, $a_x \in A_n$ and $t_x \in T_n$, but there is \emph{not} an equivalent $e_y$ where $a_y \in A_d$ and $t_y \in T_d$.

\begin{figure}
\caption{An example where the target is the dropdown button and the possible actions might be a click, mouseover, keyboard focus, or keyboard press.}
\includegraphics[width=\linewidth]{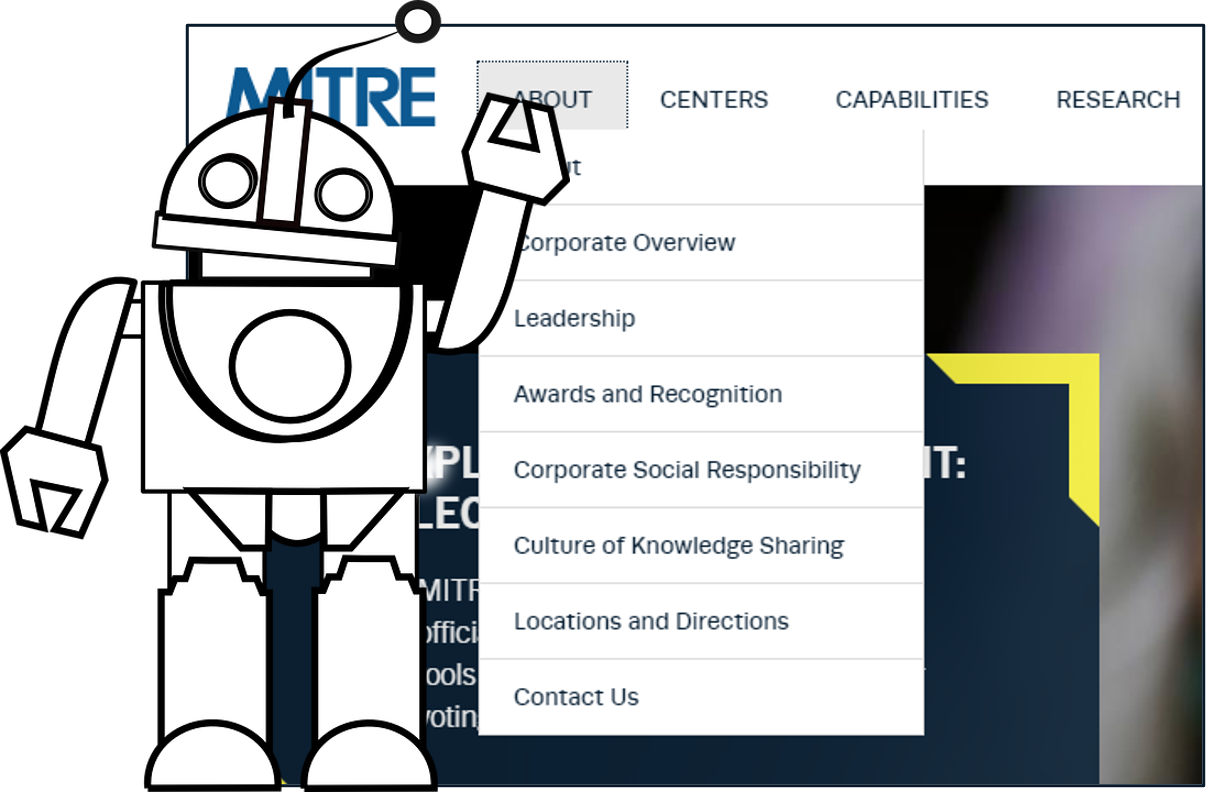}
\label{target_and_action}
\end{figure}

As a practical example, consider the robot interacting with the MITRE homepage\footnote{https://www.mitre.org/} shown in Figure \ref{target_and_action}. In this case the target is the navbar dropdown button and the possible actions might be a click, mouseover, keyboard focus, or a keyboard press. An inaccessible representation would occur if the only actions available for reaching the expanded state were a click or mouseover as a user that relies solely on a keyboard would not be able to access it. In fact, this is one of the most common examples that occurs on the real web!

A formal definition is shown in Equation \ref{access_viol}.


\begin{equation}
\begin{split}
G_{omni} &= (V_{omni}, E_{omni}) \\
G_{n} &= {V_n, E_n} \\
G_{d} &= {V_d, E_d} \\
v_{n\_start}, v_{d\_start}, v_{end} &\in V_{omni} \\
& \text{ where for some} \\
e_i, e_j &\in E_{omni} \\
\delta \left(v_{n\_start}, e_i\right) & = v_z \\
\delta \left(v_{d\_start}, e_j\right) &= v_z \\
\\
a_x \in A_n &,\; a_y \in A_d \\
t_x \in T_{v_x} \cap T_n &,\; t_y \in T_{v_y} \cap T_d \\
e_x = \{a_x, t_x\} &,\; e_y = \{a_y, t_y\} \\
\exists \left(v_x, e_x\right) \mid & \delta \left(v_{n\_start}, e_x\right) = v_{end} \\
\nexists \left(v_y, e_y\right) \mid & \delta \left(v_{d\_start}, e_y\right) = v_{end} \\
\end{split}
\label{access_viol}
\end{equation}

In Equation \ref{access_viol}, we begin by defining the three user models that are required during this comparison, the omni-user, a non-disabled user, and a user with a disability. 

Next, we define several vertices that exist within omni-user's graph $v_{n\_start}$ (non-disabled user start), $v_{d\_start}$ (user with disability start), and $v_{end}$, where $v_{n\_start}$ and $v_{d\_start}$ have outgoing edges ($e_i$ and $e_j$) that lead to $v_{end}$. Note, that it is possible for $v_{n\_start} = v_{d\_start}$ and $e_i = e_j$.

Additionally, we define edges that are capable of execution by our non-disabled user model $n$ and our user model of a user with a disability $d$. We do this by ensuring the edges use only the actions available to the given user model (e.g., $a_x \in A_n$) and targets the user is capable of interacting with (e.g., $t_x \in T_n$). We then use our definition of an edge from Equation \ref{edge_def} to construct edges available to our user models.

To determine if a representation is inaccessible for some user model with a disability $d$, we first need to determine that the state is accessible to some user model without a disability $n$. We do this by finding some vertex $v_{n\_start}$ with an outgoing edge $e_x$ that is traversable by our non-disabled user $n$. More precisely, there is some vertex $v_{n\_start}$ that contains a target $t_x \in T_{v_{n\_start}} \cap T_n$ (targets available to the user within vertex $v_{n\_start}$) that when action $a_x \in A_n$ is applied to $t_x$ triggers a traversal of the graph to $v_{end}$. We represent this mathematically by adapting the transfer function previously defined in the state machine. 

For a representation to be inaccessible, a non-disabled user must be able to access it when a user model with a disability cannot. We say a representation is inaccessible when there is no $v_{d\_start}$ with an outgoing edge $e_y$ that is activatable by our user model with a disability $d$. This failure is produced when there is no $t_y \in T_{v_{d\_start}} \cap T_d$ such that when an action $a_y \in A_d$ is applied to $t_y$ it triggers a transition to $v_{end}$. 

Finally, notice that an inaccessibility occurs as a result of a user model with a disability being unable to perform some action available to the non-disabled user ($a \in A_n$ and $a \notin A_d$) or being unable to use a target available to a non-disabled user ($t \in T_n$ and $t \notin T_d$).


\subsection{User Models}
\label{models}

As defined in Equation \ref{user_model_properties}, user models generate the $G_{omni}$ or $G_r$ graphs based on available actions and reachable states. In this proof-of-concept, we define four user models and the graphs they generate when run with Demodocus: \emph{Omni} ($G_{omni}$), \emph{Keyboard} ($G_{keyboard}$), \emph{Low-Vision} ($G_{low-vision}$), and \emph{Screen reader} ($G_{screen-reader}$). The \emph{Omni} user model is able to execute all user actions available and generates the canonical graph $G_{omni}$ for a website. The \emph{Keyboard} user model simulates a user that has limited ability to use a mouse by eliminating mouse-specific events from the available actions.  The \emph{Low-Vision} user model requires zoom and/or higher color contrast to successfully navigate a page. The \emph{Screen Reader} user model simulates a user that requires proper labeling to include an element in its set of targets (i.e., it must be able to understand a target in order to interact with it). In Table \ref{user_model_table} we show examples for how various user types may be restricted based on their simulated disability.

\begin{table}[htb]
\begin{tabular}{p{.2\linewidth}|p{.32\linewidth}|p{.32\linewidth}}
User Model & Possible Actions $A_r$ & Possible Targets $T_r$  \\
\hline
\hline
Omni       & All actions & All targets \\
\hline
Keyboard   & Keyboard events &  Focusable and \break navigable  targets \\
\hline
Low Vision   & All actions & Targets with \break sufficient size \break and contrast \\
\hline
Screen Reader   & Keyboard events & Labeled targets \break that exist in \break the accessibility tree  
\end{tabular}
\caption{{Examples for how a user model's restriction are reflected in the sets of action and targets available to them. }}
\label{user_model_table}
\end{table}

Our prototype implemented these user models using a three step Perceive, Navigate, and Act approach. In this approach, the simulated users were required to be able to perceive a target existed, be able to navigate to the target, and then trigger an action on the target.  

The \emph{Perceive} ability for each user model defines how the user model interprets the web page. For the \emph{Keyboard} user model, this may be all elements visually present on the page; for other user models (e.g., \emph{Screen-reader}, \emph{Low-vision}), it may mean access only to information presented by the assistive technology or visible through a smaller viewport. This step can act as a filter for the remaining abilities; if a user cannot perceive an element, it is unlikely they would navigate to and act on it. 

The \emph{Navigate} ability defines how a user navigates to an interactive element within a page. As an example, for a \emph{Keyboard} user model this primarily entails tabbing, but could also include keyboard shortcuts, especially those given by screen readers in the case of the \emph{Screen-reader} user model. 

Finally, the \emph{Act} ability defines how a user model is able to interact with a page, and often is as simple as defining a set of JavaScript event types that the model is allowed to use. For the \emph{Keyboard} user model, this would mean restricting the set of JavaScript events to those only created by the keyboard such as \texttt{keyup} or \texttt{keydown}.

\begin{figure}
\begin{subfigure}{.5\textwidth}
  \centering
  \includegraphics[width=0.8\linewidth]{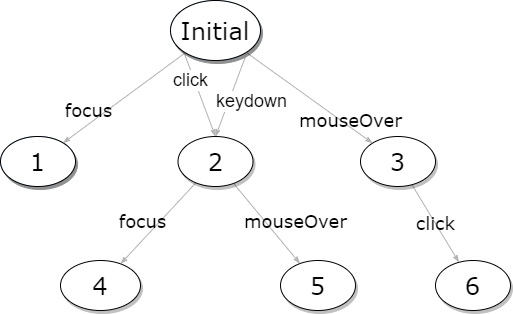}
  \caption{The graph generated by simulating a user without a disability.}
  \label{fig:omniuser_graph}
\end{subfigure}%
\\
\begin{subfigure}{0.5\textwidth}
  \centering
  \includegraphics[width=0.8\linewidth]{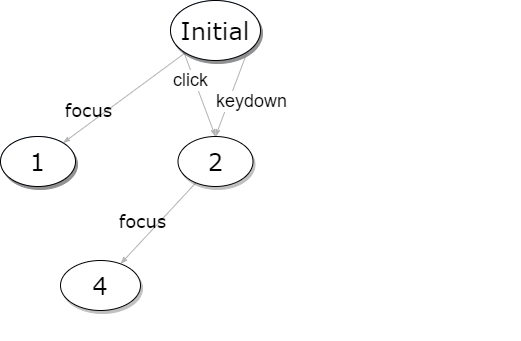}
  \caption{The graph generated by simulating a user with a disability. States 3, 5, and 6 are not included.}
  \label{fig:keyboard_user_graph}
\end{subfigure}
\\
\begin{subfigure}{.5\textwidth}
  \centering
  \includegraphics[width=0.8\linewidth]{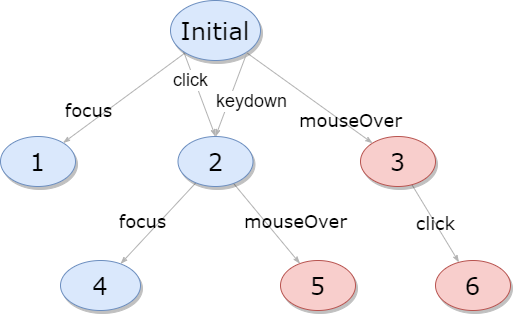}
  \caption{Visualizes the difference between the graph generated between the users with and without disabilities. States 3, 5, and 6 are all shown to be reachable by a user without a disability but not reachable by a user with a disability.}
  \label{fig:graph_differential}
\end{subfigure}
\caption{Shows the process of comparing the generated graph of users with and without disabilities to find accessibility violations.}
\label{fig:find_violations}
\end{figure}

Figure \ref{fig:find_violations} shows an example of the process combining the user models with the graph exploration methods to evaluate a web page for accessibility. We start by exploring the states of the web page reachable to a user without a disability as shown in Figure \ref{fig:omniuser_graph}. In this case, we generated 7 different states, including the initial state, using various events such as \emph{onFocus}, \emph{onClick}, and \emph{onKeydown}. We then crawled the web page by simulating a user with a disability that could only use the keyboard; the resulting graph is shown in Figure \ref{fig:keyboard_user_graph}. Finally, we compare the generated graphs to determine which states are unreachable for our user with a disability. As Figure \ref{fig:graph_differential} shows, states 3, 5, and 6 are unreachable and thus candidates for violations. Analyzing further, we can speculate that the \emph{onMouseOver} edges need a keyboard alternative.

\subsubsection{Ease of Use Scoring}

The approach described in section 3.2 allows Demodocus the opportunity to determine how difficult reaching any particular state might be. Note that a violation found by any of these user models would be reported as a potential accessibility violation for the page as a whole. However, instead of focusing purely on violations we can extend the analysis to also include usability analysis for users with disabilities. Not only can we determine if an application is accessible but can lend insight into whether or not it is usable for users. A commonly-seen example of a webpage that is accessible but not usable is one with a poor implementation of bypass blocks. To reach the main content, a user may be required to do a significant amount of tabbing; this is technically accessible, but practically unusable.

We begin by creating a scoring function, \emph{f(x)}, for each part of the Perceive, Navigate, and Act interaction approach. The scoring function takes as input some metric relating to the ability type and outputs a decimal value between 0 and 1. A measurement of 1 represents a completely accessible and effortless experience, while 0 represents a completely inaccessible or unusable activity. 

As an example, consider a \emph{Keyboard} user modeled to have perfect vision (Perceive), requires tabbing to elements (Navigate), and uses only keyboard events (Act). The Perceive scoring function might score a 1 if the target is plainly visible or a 0 otherwise. We could then set the Navigate function to be inversely proportional to the number of tabs required to reach the target. Once the target is reached, we will then perform a check against whether the target supports any keyboard events, scored 1 for yes and 0 for no.

\begin{figure}
\caption{A weighted graph demonstrating the ease or difficulty of reaching each state. States with a score of 0 are unreachable, while a score of 1 denotes a perfectly accessible and usable experience.}
\includegraphics[width=1\linewidth]{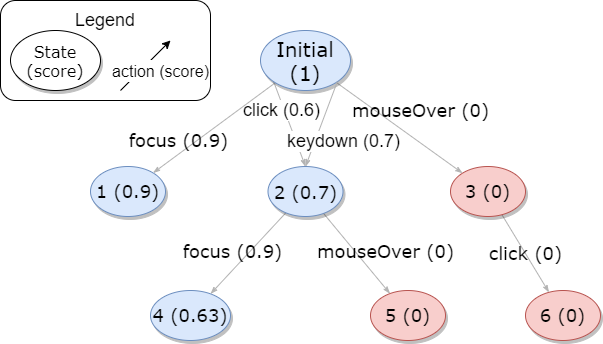}
\label{fig:weighted_graph}
\end{figure}

We score the state transition by multiplying the values from each step (e.g., for the example above 1 $\times$ 1/number of tabs $\times$ 1, though this overly penalizes tabbing). Note that a 0 on any step of the process will result in a score of 0 for the overall state transition, indicating the inability of the user to traverse some edge. As a result of scoring the edges, we now end up with a weighted graph. Since any workflow can be represented by a path through the weighted graph, we can determine the score of the workflow by multiplying the edges on the path; the final score will determine the overall usability of the task! Figure \ref{fig:weighted_graph} shows the graph from Figure \ref{fig:graph_differential} with the weights included along the vertices and edges. 

While scoring for usability was not a focus of our research, we discuss it here to emphasize to readers that the Demodocus framework allows for this type of analysis. This has been a long-sought metric for web applications that is unattainable with the statelessness of current automated approaches.

%

\subsection{State Equivalence}
\label{state}

Due to our specific application of measuring accessibility, we have had to adapt the foundational methods described in the Dincturk Hypercube model \cite{dincturkAjax} and our prior JavaScript crawling work \cite{jbrunelleDissertation}. Specifically, we have extended the prior state equivalency algorithms to more accurately reflect the user experience (i.e., determine what states an average user may consider equivalent). To achieve this goal, we implemented and evaluated two distinct state comparisons techniques: A three stage approach that includes increasingly lenient string and DOM comparisons, and a templating approach to determine missing or changed DOM elements aggregated over time. In this section we describe the algorithms and discuss the performance of each approach.

\subsubsection{Three-Stage Comparator Pipeline}

The first stage of the state comparator is a simple HTML string comparison. This is the strictest comparator in our approach. If we find the strings to be equivalent, then we know with the highest probability that the states are equivalent. We cannot be certain, as there may be additional information contained in cookies or on the server that has changed and that we are not inspecting. While this approach gives us an easy way to determine exact state equivalence, it is unlikely to find states that a user would consider matching. Consider that a simple timestamp on a page would break this comparator. If the states are equivalent with respect to this comparator, the state equivalency process will return that the states are equivalent and proceed no further. However, if the states fail this comparison (i.e., their HTML strings do not match exactly), we move on to the DOM comparator.

The DOM comparator is the second stage in our tiered comparison approach. This comparator defines state equivalency as having the same DOM structure and is primarily concerned with whether nodes have been added to or removed from the DOM. If nodes have been added or removed, the DOM comparator will judge the state to have changed and will return that the states are different and exit the state equivalency process. Otherwise, the DOM comparator is unable to find significant evidence that the states are equal or equivalent and will proceed to the final stage of the comparison process. 

Note that the exit criteria are different in between the exact HTML string and DOM comparators. Whereas the HTML string comparison exits on finding state equivalency, the DOM comparator does the opposite; exiting only after determining that the states are different.

The final stage of our original comparator pipeline is a fuzzy text comparator. The comparator calculates the Levenshtein distance between the contents of the text nodes in each of the two states and considers them equal if the distance is less than some threshold. This comparator is required, in addition to the previous two, due to the fact that our use case in determining accessibility requires text changes as a possible accessibility feature. For instance, content on a page may change upon reload (e.g., due to a live feed). This change will not be caught by the first two comparators. The exact string comparison will pass the states downstream, as the raw HTML text has been modified and the DOM comparator will fail since the structure of the DOM has remain unchanged despite the added text. The threshold for considering states equal does require tuning, and in some cases may be a function of the page size or several other factors. 

\subsubsection{Page Templating}

While the Three-Staged approach described above works for simplistic pages, it struggles when any amount of uncertainty is added to the page, such as text changing upon page reload or small element location changes. While the fuzzy string comparison allows for changing, we found the approach limited in its requirement for tuning and its inability to scale to more complex applications. Our motivating insight for the page templating approach described below is that much of the changing content that is not a result of user interaction changes consistently and follows a similar structure to changes seen in the past. For example, a message board or news site may have constantly rotating information, but the information is likely to be presented in a similar format during each time step.

The page templating approach begins immediately upon reaching the initial state. Before running the state exploration process, we instead build up an expected template for the design and structure of the web page by determining what content changes automatically inside the page (e.g., rolling news feed) and what changes between page reloads (e.g., new message posts). Any content that is found to be changing after the page stabilizes (i.e., fully loaded and finish any loading animations) or between page reloads is marked as unstable. Labeling a section as unstable denotes that content changes may not be due to user input and thus it should be ignored for the purposes of state deduplication. We repeat this process several times, reloading the initial state of the page, recording each instance of the page and its automatically changing content, and then saving it into a time map. After completing this process, we build the canonical template of the page by aggregating across each instance captured in the time map and marking all sections that have been found to be unstable. 

While crawling, new states are declared when the changes between the pages occur outside of any unstable regions (i.e., unstable regions are factored out of the comparison). If a state is determined to be a unique representation, then its content is added to the previous state's template to determine the template of the current state. In this way, the template description of state content flows logically through the graph. While not an infallible algorithm, its introduction expanded our capability to logically deduplicate states in our graph.

\section{Future Work}

We consider the Demodocus framework to introduce a novel testing strategy for the automated accessibility testing of web applications. As such, we consider there to be many areas for further expansion and research: extending user models, adapting the framework for usability, and using statistical and machine learning approaches for determining state equality.

In this paper we introduced several possible user models, but as the reader may have noticed, they are significantly simplified in order to reduce the overall complexity of the framework. The increased fidelity of the existing user models (e.g., integration of information scavenging theory) and the introduction of new user models (e.g., users with cognitive disabilities) would significantly benefit our approach, allowing us to better capture user experiences and accessibility violations. 

While discussing user models in section \ref{models}, we briefly mentioned how the graph generation and simulation approach may be used to determine the user experience. However, the scoring functions currently implemented are naive and do not capture the intricacies of a user's full interaction paradigm. For instance, the current approach significantly penalizes long or multi-step processes, but these may not necessarily be poorly usable; in fact, perhaps the creators have broken a complex subject down into digestible chunks.

A constant problem for any approaches relying on state definition is the task of accurately reducing the state space to a manageable level. We discussed our state equivalence model, but even with significant effort failure cases are relatively easy to find. For example, an infinite scroll list where more content is continually added to the page would require additional intelligence to diagnose. We focused primarily on deterministic measures for this opening study, but believe statistical and machine learning approaches should be attempted for state deduplication as they will likely have greater state adaptability. 

To promote further research, we have open sourced the Demodocus framework prototype \cite{demodocus_github} so that it is freely available for download and extension. We hope readers will use this code as the basis for their own automated accessibility tools and will continue to push the state of the art both for the topics mentioned in this section and the many others that can push this technology forward.

\section{Conclusions}

In this paper we presented the Demodocus framework which describes a novel testing framework for automatically finding potential accessibility violations in dynamic web applications. Current automated accessibility testing tools are stateless, testing only on the specific state of the page as when they were activated. While these tools perform incredibly well, they only test at most 50\% of known accessibility issues and scale poorly with the growing interactivity of the modern web. 

 In comparison, the Demodocus framework uses a web crawler capable of generating a graph of all states available in the page. Using the generated graph, we are then able to determine the reachability of various states. States that are reachable by non-disabled users but not users with disabilities denote the existence of accessibility violations and are reported to the testers.
 
 We firmly believe the capability to automatically evaluate interactive web pages will allow government and businesses to achieve accessibility compliance with significantly less effort and expenditure while also increasing the quality of their applications. \newline
 
\noindent\rule{\linewidth}{1pt}

\noindent \copyright 2021 The MITRE Corporation. ALL RIGHTS RESERVED.

\noindent Approved for Public Release; Distribution Unlimited. Case Number {21-3015}.


\bibliographystyle{abbrv}
\bibliography{_mybibtex}

\end{document}